# Sequential phase-locked optical gating of free electrons

Fatemeh Chahshouri[1,*], Nahid Talebi[1,2,*]

[1]*Institute of Experimental and Applied Physics, Kiel University, 24098 Kiel, Germany*
[2]*Kiel, Nano, Surface, and Interface Science − KiNSIS, Kiel University, 24098 Kiel, Germany*

E-Mail: talebi@physik.uni-kiel.de; Chahshuri@physik.uni-kiel.de

**Abstract**

Recent progress in coherent quantum interactions between free-electron pulses and laser-induced near-field light have revolutionized electron wavepacket shaping. Building on these advancements, we numerically explore the potential of sequential interactions between slow electrons and localized dipolar plasmons in a sequential phase-locked interaction scheme. Taking advantage of the prolonged interaction time between slow electrons and optical near-fields, we aim to explore the effect of plasmon dynamics on the free-electron wavepacket modulation. Our results demonstrate that the initial optical phase of the localized dipolar plasmon at the starting point of the interaction, along with the phase offset between the interaction zones, can serve as control parameters in manipulating the transverse and longitudinal recoil of the electron wavefunction. Moreover, it is shown that the polarization state of light is an additional control knop for tailoring the longitudinal and transverse recoils. We show that a sequential phase-locking method can be employed to precisely manipulate the longitudinal and transverse recoil of the electron wavepacket, leading to selective acceleration or deceleration of the electron energy along specific diffraction angles. These findings have important implications for the development of novel techniques for ultrafast electron-light interferometry, shaping the electron wave packet, and quantum information processing.

**Introduction**

The emergence of photon-induced near-field electron microscopy (PINEM) in 2009[1] and its theoretical quantum description[2–4] have revolutionized the field of light-matter interaction. Producing femtosecond electron pulses by the combination of ultrafast laser pulses and electron microscopes to probe the ultrafast charge oscillations and the dynamics of excited near-fields have provided a powerful tool for electron holography and phase retrieval[5], as well as controlling the shape of free-electron wavepackets[6] and constructing attosecond electron pulse trains[6–9]. This breakthrough not only extends new prospects for quantum information processing[10] but also facilitates shaping of the electron beam[11–14].

During this interaction, overcoming the condition for energy-momentum conservation leads to quantized energy exchange between light and the electron that leads to the momentum modulations predominantly along the longitudinal direction[13]. The unavoidable transversal momentum recoil[15–17] leads to transversal momentum modulations as well. Consequently, in addition to the coherent bunching of the electron wave function in energy space and Rabi oscillations[4] among quantum states separated by multiples of the photon energy[4], the periodic Lorentz force[15] acting on the electron by localized plasmons[15] serves as a phase and amplitude grating for elastic diffraction[18]. Therefore, the optical near-field, is responsible for transferring energy and momentum to free electrons, and can control longitudinal[13,15,16,19] and transverse recoil[13,15] of traveling electron beams.

The free-electron and light interactions in PINEM experiments could be tailored to cover the full range of weak and strong interactions. The confined near-field modal volumes in microcavities[20] and highly localized plasmonic mode[21], can overcome the weak phase-matching problem. Another approach to achieve higher energy-momentum matching is precisely matching the phase velocity of the light wave and the group velocity of the electron wave function[22]. Recent experiments, which use the inverse Cherenkov effect[23] or whispering gallery modes[20] have shown resonant phase-matching and an exchange of hundreds of photon quanta with a single electron over long interaction distances. Another efficient way to satisfy phase-matching conditions is to take advantage of slow electrons[15,24]. The prolonged interaction time between low-energy electrons and light waves increases the possibility of mapping the dynamics of several near-field oscillations[15] between probe electrons and spatially localized light-field of small nanostructures.

Careful control of the phase modulation of slow electrons via the polarization[13] and the spatial profile of the optical near-field beyond the adiabatic approximation[24,25] has opened the way to manipulate free electrons and control its longitudinal inelastic energy transfer, as well as its transverse elastic recoil[26]. This results in selective control of the electron energy and diffraction angles, enabling the implementation of clever acceleration/deceleration mechanism within an arbitrary angular deflection range. Pumping samples with two-color laser pulses[6] and having two spatially separated near-field zones[6,27] are other ways to control the final longitudinal energy transfer and quantum-phase modulation of the electron wavefunction. Electrons flying through two spatially separated near-field lights can carry information about the amplitude and phase of each optical field separately. The first interaction alters the amplitude and phase of the electron wavepacket. There, the already shaped electron wavepacket interacts differently with the near-field light at the second interaction zone, compared to an unmodulated electron pulse[5].

Here, we aim to map the impact of the phase oscillation of the dipolar plasmons in a double interaction phase-locked system to control the longitudinal and transversal distribution of the slow electron wavepackets. We numerically show that the initial phase of the optical cycle and the phase offset between the localized dipolar plasmon can control the shape and diffraction angle of the electron. Particularly the initial phase of the light at the first interaction point, strongly affects the final shape of the electron wavepacket, mainly due to the direction of the wiggling motion exerted on the electron in the first interaction zone. We define the polarization dependence excited modes as a quantity for selectable enhancement or cancellation of specific momentum orders. We further demonstrate that depending on the spatial profile of the dipolar modes and relative phase between two near fields, the exerted transverse electromagnetic force can lead to the deflection of electrons within the second interaction zone.

**Results and discussion**

The discussion above has showcased that the interaction between electron wave packets and laser-induced plasmon excitation yields amplitude and phase modulations in the electron wave packet. To provide a movie-like access to such dynamics, we used our self-consistent Maxwell-Schrödinger numerical toolbox[16,25]. In this process the electron wavepacket ($\psi(\vec{r}, t)$) evolution versus time is calculated by solving the Schrödinger equations using the minimal-coupling Hamiltonian[15] in the vicinity of a laser-induced gold nanorod. The properties of the plasmonic near-field in each time step is calculated based on the finite-difference time-domain (FDTD) method, where the permittivity of the gold was modeled using a Drude model complemented by two critical point functions[16], and then interpolated into the Schrödinger frame. After the interaction is completed in the Schrödinger frame, the final electron wave packet is used to calculate the energy modulation and electron recoil. Then the expectation value of the electron kinetic-energy operator is calculated as follows:

$$\langle \psi(x,y;t \to \infty)|\hat{H}|\psi(x,y;t \to \infty)\rangle = \frac{\hbar^2}{2m_0} \iint dk_x dk_y (k_x^2 + k_y^2) |\tilde{\psi}(k_x, k_y; t \to \infty)|^2, \quad (1)$$

where $(x, y)$ and $(k_x, k_y)$ in Eq. (1) denote the real and reciprocal space coordinates, and $\tilde{\psi}(k_x, k_y; t \to \infty)$ is the Fourier transform of the wave function after the interaction. By assigning the kinetic energy of the electron as $E = \hbar^2(k_x^2 + k_y^2)/2m_0$, and the scattering angle as $\varphi = \tan^{-1}(k_x/k_y)$, the inelastic scattering cross section $\sigma(E, \varphi) = (m_0/\hbar^2)|\tilde{\psi}(E, \varphi; t \to \infty)|^2$ for different inclination angles is obtained, where $m_0$ is the electron mass, and $\hbar$ is the reduced Planck's constant[13]. We define the inelastic scattering cross section value as a quantity for distinguishing unique features that can be measured due to strong interaction.

The fact that the electron is energetically bunched in discrete photon order is delicately explained via a one-dimensional electron model that neglects the recoil experienced by the electron[2]. The discrete probabilities of each spectral peak associated with the exchange of $n$ quanta of energy between the electron and scattered field are given by expanding the wave function by the Bessel series using Jacobi-Anger relation[15]:

$$\psi(x,t) = \exp[(ik_x^{el}x - \Omega t)] \sum_n i^n J_n(|g|) \exp\left[in\left(\omega_{ph}/v_{el}\right)x - n\omega_{ph}t + n\sphericalangle g\right], \quad (2)$$

where $J_n$ is the $n$th Bessel function of the first kind, and $g$ is the strength of the electron–light interaction $g = (e/\hbar\omega_{ph}) \int_{-\infty}^{\infty} dx' \tilde{E}_x(x', y) e^{-ix'\omega_{ph}/v_e}$. $\tilde{E}_x$ is the Fourier transform of the optical electric field component along the $x$ direction, and $v_e$ is electron velocity, respectively. $k_x^{el}$ is the initial longitudinal center wavenumber of the electron wavepacket, and $\omega_{ph}$ is the incident photon angular frequency.

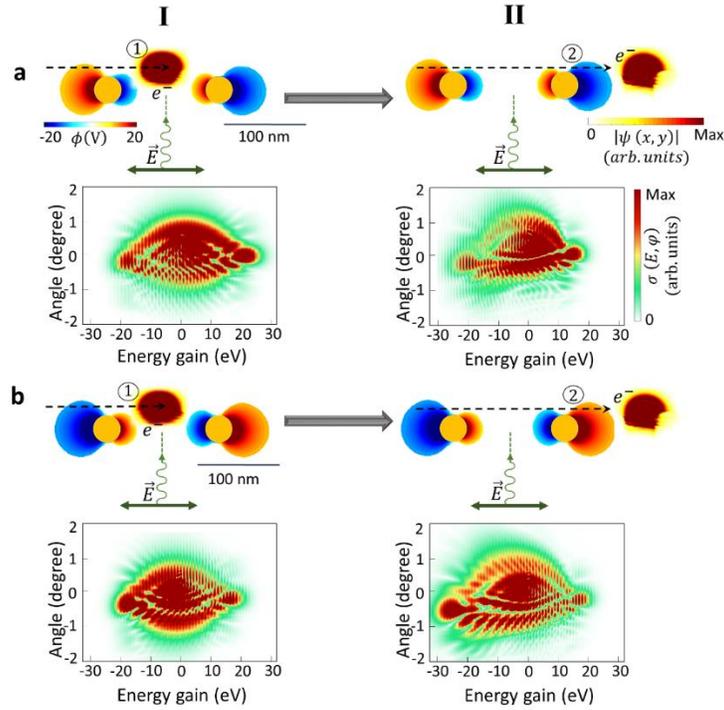

**Figure. 1. Sequential free-electron and photon interactions.** Electron modulation spectra after interacting with near-field of (I) first gold nanorod (1), and (II) second nanorod (2). Final modulation of the amplitude of the electron wavepacket in real space (top schematics) after the interaction with the dipolar near-field modes of each nanorod and its inelastic scattering cross-section (bottom schematics). The phase of the optical near field of each nanorod at the start of the interaction with the electron is the same in the dual interaction scenario (as depicted by $\text{Re}\{\phi(V)\}$ at a depicted time in the figure, where $\phi(V)$ is the scalar potential). The initial phase of the near-field light is reversed in (b) compared to (a). The initial center kinetic energy of the electron beam is 600 eV, and its longitudinal and transverse broadenings are 56 nm and 36 nm, respectively (FWHM). The center wavelength of the laser pulse is 700 nm, its FWHM temporal broadening is 18 fs, and its peak field amplitude is 1 $\text{GVm}^{-1}$. Dashed arrows show the trajectory of the electron.

In this work, the coherent control of the electron wave function is mediated by two spatially separated localized plasmonic fields. Here, we study the importance of the initial phase of the optical mode in a system involving a slow electron wavepacket passing through the near-fields of phase-locked plasmonic gold nanorods. The low velocity of the electron inherently increases the interaction time and, thus, for a given local field amplitude, enhances the experienced electron recoil. We considered the interaction of a Gaussian electron wave packet at the kinetic energy of 600 eV, with 56 nm longitudinal ($W_L$), and 36 nm transverse ($W_T$) broadenings excited by a linearly x-polarized laser pulse with the center wavelength of $\lambda = 700$ nm and temporal broadening of 18 fs. The gold nanorods responsible for transferring momentum to electrons are placed at 100 nm distance from each other. We considered nanorods with a radius ($r = 15$ nm) capable of hosting only dipolar modes. Simulations for the scattering of the electron from one isolated nanorod repeat the same results as the shape of the electron after the first interaction zone in the sequential scheme with 100 nm spacing (see supplementary Figure. 1). Due to the large spacing between the nanorods, the hybridization of plasmonic modes do not occur and only a weak radiative interaction between rods slightly alters the plasmonic resonances of the double-rod system compared to a single isolated rod (Figure. 1). Figure. 1, shows electron modulation in the real and the momentum representation after experiencing each step of the dual near-field interaction. The top schematic in each section visualizes the dipolar mode when the electron starts interacting with each nanorod, and finally electron modulation in the spatial domain. After the interaction is finished, a train of energy combs displaced at exactly the incident light-field energy is observed in the electron energy spectra (Figure. 1a and b; lower rows). The parameters are configured here to match the synchronicity condition $\lambda_{ph}v_{el} = 2rc$ [15], in order to have a synchronous dipolar field oscillation with its period being matched to the propagation time of the electron beam within the effective interaction length. $\lambda_{ph}$ here is the wavelength of the plasmonic resonances.

Generally, a spread of the electron momentum distribution along both longitudinal and transverse extend is observed. The longitudinal electric field components of the near-fields provide momenta to bridge the phase mismatch for energy transfer to the electron, in fact the localized near-field couples to the electron at the initial momentum state $p_e = \hbar k_e$. As the electron absorbs and emits quanta of photon $n$, its wavefunction evolves into a superposition of momentum distribution $p_e = \hbar\left(k_e + n\left(\omega_{ph}/v_e\right)\right)$. Consequently, an energy comb with the spacing between the peaks ascertained by the exciting photon energy ($\hbar\omega_{ph}$) is formed[2]. On the other hand, the transverse field component induces sideways diffraction of the electrons. The arrangement of diffraction orders at different energies depends on the electron velocity, the optical near-field momentum distributions, and the topology of nanoparticles[15]. Selection rules for the diffraction peaks along $E = 0$ axis in the vicinity of the plasmon of gold nanorod are displaced by orders specified by $\delta k_y^{el} = 2|k_y^{ph}| = 2\sqrt{k_{x,c}^2 - k_0^2}$, where $k_y^{ph}$ represents the damping ratio of the evanescent tail of plasmons along the y-axis, and $k_0$ is the photon wave number in free space. Hence the occurred diffraction process is a sequential two-photon process.

Here we have employed two scenarios to investigate the impact of the initial phase of the oscillating dipolar mode on the electron modulation. The first configuration (Figure. 1a,) involves the electron experiencing recoil influenced by the positive phase of the dipolar mode as its initial interaction condition. In the second scheme (Figure. 1b), the electron recoil is controlled by the initially negative phase of the dipolar mode in both nanorods. Following each interaction process, the inelastic scattering cross-section map illustrates the influence of synchronization between the arrival of the electron wave packet and the phase of the near-field light in a dual interaction system.

As depicted in Figure. 1a, after the electron passes through the first near-field zone (I), only a slight asymmetry in the intensities of momentum orders is observed in the overall inelastic scattering cross-section. The oscillating *x*-polarized dipolar near-field induces the rotational restoring force acting on the electron, and as the result of the back-and-forth force on the electron, a unified transverse recoil is observed across energy distribution. Therefore, the electron wavepacket experiences a total transverse recoil toward both the positive and negative y-axis $-1° \leq \varphi \leq +1$, together with longitudinal inelastic energy exchange within the range of $-20\text{eV} \leq E \leq 20$ eV. This energy comb reveals distinct sidebands for both positive and negative longitudinal momentums, which indicates energy loss and gain (acceleration/deceleration) processes on the electron pulse during the interaction. Subsequently, as the already modulated electron continues to interact with the in-phase rotational force of the second near-field (II), some of the previously populated momentum ladders in the electron energy gain are depopulated, therefore we can see reduction in the level of momentum occupancy along both vertical recoil and horizontal energy exchange order for loss channels. In contrast, an increase in the probability of occupying a higher energy level and an expansion in the diffraction angle distribution for the gained-energy electrons are observed.

This scenario is different for a system involving first interactions with a negative phase of the dipolar mode. As shown in Figure. 1b, after a sequential interaction, we can see numerous fine interference features of discretized energy and recoil states similar to the first case. However, there is a distinction in the visibility of the transverse momentum levels. Here, the distribution of the electron wavepacket along the longitudinal and transverse directions shows discretized momentum levels in the negative range of spectrum (energy loss region).

Here, we indicate a robust correlation between the observed electron modulation and the initial phase of the harmonic evanescent field distributions at the start of the interaction. The quantum phase modulation, governed by the phase-controlled double interactions is better visualized in inelastic longitudinal PINEM spectrum (supplementary Figure. 2) which shows phase dependent selectable population and depopulations in the final electron energy spectra. Where the transversal differential scattering cross section (supplementary Figure. 3) demonstrates the effect of the arrival time of the electron wavepacket to the near-fields in a dual interaction scheme on hampering and enhancing diffraction recoil. We attribute this asymmetry to the direction of the wiggling motion exerted on the electron via the interaction with the in-phase oscillating field. The acting Lorentz Force on the electron wavepacket, initiates this motion and leads to the dynamical deflection of the electron in the near-field zone, therefore different parts of the electron wavepacket are exposed to different coupling strengths between the field and the electron wavepackets. The acting Lorentz Force on the electron wavepacket, initiates this motion and leads to the dynamical deviation of the electron in the near-field zone from its initial shape and trajectory. As a consequence, distinct sections of the electron wavepacket experience varying degrees of interaction strength between the electromagnetic field and the electron wavepacket (see supplementary Movie 1).

To further enhance our understanding of the impact of the initial phase of the optical mode on the electron modulation in a Ramsey-type interaction, we will explore the influence of the polarization of the laser on the electron energy exchange and the experienced recoil. Upon excitation with the linearly polarized laser a dipole is induced in the parallel direction to the electric field of the laser radiation. For an inclined excitation this will lead to an inclined Lorentz force acting on the electron with respect to the electron trajectory[13] that initiates an inclined wiggling motion as well[13].

Moreover, due to the asymmetric projected field along the electron propagation direction, the electron distribution in the momentum representation rotates toward specific diffraction angles and energy ranges. In the following, we investigate the interaction of an electron wavepacket with the oblique dipolar oscillation of two phase-locked gold nanorods with zero relative phases between them. The simulation parameters are chosen similarly to those presented in Figure. 1, but the direction of the laser illumination changed from $\theta = 0°$ to $\theta = -30°$. Angular deflections with complex diffraction patterns in each photon absorption and emission order are observed (Figure. 2).

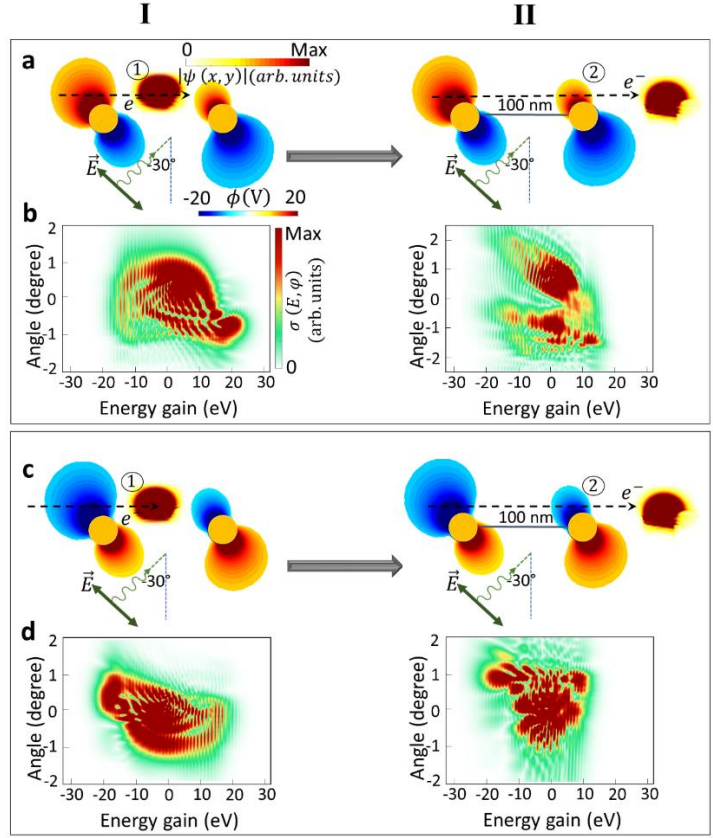

**Figure. 2. Polarization-dependent phase-locked electron-photon interaction.** (I) Single-interaction and (II) double-interaction scenarios. The amplitude of the electron wavepacket after propagating through the near-field zone excited by a linearly polarized light at the incidence angle of $-30°$ in the (a) real- and (b) momentum-space representationsThe electron begins its interaction with nanorods when the plasmonic dipolar field has either a (a, b) positive initial phase (upper box)or (c, d) a negative initial phase (lower box). The laser pulse has a center wavelength of 700 nm, electric field amplitude of $E_0 = 1 \text{ GVm}^{-1}$, and temporal FWHM broadening of 18 fs.

Already after the first interaction region, the effect of the initial phase of the near-field distribution and its asymmetry with respect to the electron trajectory becomes apparent. In fact, these aspects lead to an asymmetry of the electron wavepacket distribution in the momentum space for both cases, particularly along the transverse direction (compare Figure. 2b to 2d). Travelling to the next near-field zone, the wiggling motion of the plasmonic field induces rotational oscillations in the momentum space of the already shaped electron wavepackets. Therefore, the modulated electron experiences either a repelling or attracting force toward the nanostructure, and it leads to the population of higher momentum orders or depopulating to a lower momentum order. The critical parameters for controlling the quantum path interferences are the distribution of the momentum states after the first interaction and the direction of the induced near-field dipole with respect to initial electron trajectory. As a result of interaction with the dipolar resonances of both nanorods, the transversal diffraction order is split into two regions with $\varphi \leq 0$ and $\varphi \geq 0$, while the energy spans over a range of $-20\text{eV} \leq E \leq 20$ eV. The transversal electromagnetic force in sequential in-phase interactions exerts a significant elastic diffraction obvious from the momentum space distribution of the electron wavepacket and interestingly, a substantial upwards deflection of the electron in the real space only after a few nanometer away from the interaction zone. The later happens due to a stronger diffracted intensities towards the +y-axis.

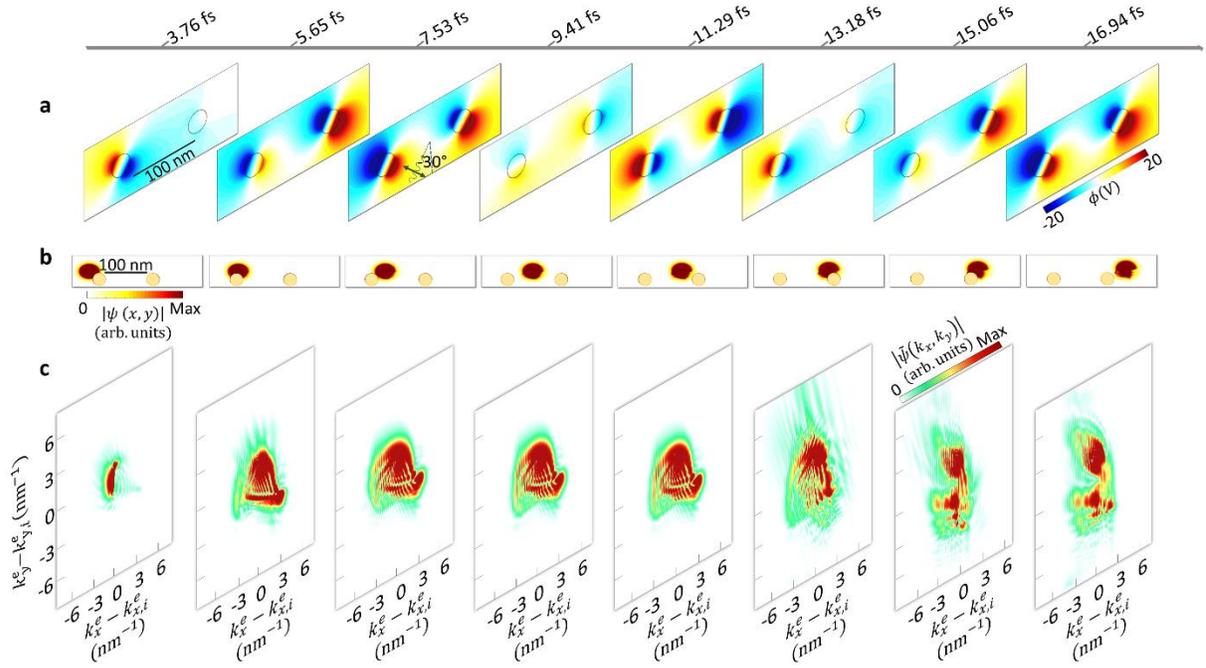

**Figure. 3. Transversal and the longitudinal dynamics of the evolution of the free electron wavepackets during its interaction with localized plasmons excited by an obliquely polarized laser ($\theta = -30°$).** Parameters of the laser pulse are $E_0 = 1$ GVm$^{-1}$ (field amplitude), $\lambda =700$ nm (center wavelength), and $\Delta\tau =18$ fs (FWHM temporal broadening). The electron wavepacket has an initial kinetic energy of 600 eV, with 56 nm and 36 nm longitudinal and transverse broadenings, respectively (FWHM). (a) Spatial profile of the scaler potential at depicted time steps. (b) The amplitude of the real-space and (c) momentum-space electron wavepacket distributions at the corresponding time steps.

To better elaborate on the sequential phase-locked electron-photon interaction of the electron wavepacket with two rods, the dynamics of the interaction for the case with a positive initial phase are shown in Figure. 3. Within the time frame of 0 to 9.41 fs, as the electron passes (Figure. 3b) by the neighborhood of the first nanorod, the oscillating dipolar localized plasmon (Figure. 3a) leads to a Lorentz force and results in a circular motion of the electron in the momentum representation (Figure. 3c). Then, the electron starts being populated in the momentum space along both the x and y axis. As the electron moves to free space between nanostructures, its momentum distribution remains constant, revealing that the electron cannot interact with the free-space electromagnetic radiation. Arriving to the second oscillating near-field, the electron wavepacket experiences a reshaping in the momentum distribution along both x and y directions. Thereby, the second near-field distribution evacuate some of the already occupied momentum states and populated new momentum orders. The overall experienced phase by the electron over multiple cycles of the light field (three oscillations for each near-field) and the direction of field oscillation (that controls the direction of the wiggling motion) affect the final span of the wavepacket in the momentum space, along both the transverse and longitudinal directions. The nonzero asymmetric force asserted by the oscillating fields cause significant electron deflection in the transverse direction after interacting with the second field, and more interestingly a strong diffraction into two momentum orders as large as 330 $k_0$ are observed. Compared to the free-space Kapitza-Dirac effect [28,29], where the electron could be occupied to the momentum orders of only $2k_0$, near-field mediated sequential control of the electron could open fascinating opportunities for routing the motions of electrons for electron-wave interferometry applications.

Simulation of dual interaction system for longer electron wave packet along the longitudinal direction (refer to supplementary Figure. 4) yields results consistent with those demonstrated in Figure. 3. Some differences arise from the shorter effective interaction length for a smaller electron wavepacket. On the other hand, for the broader electron (longitudinal broadening of $W_L = 120$ nm), the electron can experience six dipolar oscillations per nanorod and have more effective interaction length. This results in well-separated momentum ladders and more pronounced real-space deflection.

Controlling the outcome of the random walk by a few parameters, such as the laser polarization and gap distance between the nanorods, in a Ramsey-type experiment makes the proposed approach an efficient technique for shaping the electron wavepackets by encoding the roles of phase-locked oscillations into the energy transfer and transverse momentum modulation. Moreover, our scheme benefits from the co-excitation of both rods with the same laser beam.

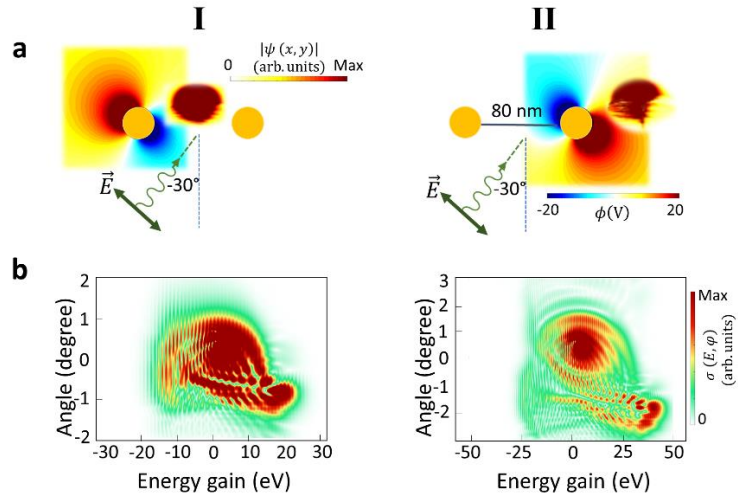

**Figure. 4. Phase controlled sequential interactions with out-of-phase optical cycles of near-fields.** The electron wavepacket at 600 eV initial kinetic energy, excited by an inclined linearly- polarized laser pulse at the wavelength of 700 nm, and with the field amplitude of $E_0 = 1$ GVm$^{-1}$ represented in (a) real space, and (b) momentum space, after the (I) first interaction and (II) second interaction zones.

It has been previously demonstrated that in the Ramsey-type experiment, the distribution of the final energy spectrum[27] is influenced by the relative phase between two interaction zone[27]. In-phase fields can enhance the final interference features in a constructive manner and therefore the extend of the PINEM pattern, whereas the two fields with opposite phases can cancel energy exchange. Merging the findings from the Ramsey experiment with the effect of the transverse Lorentz force opens a new dictionary to control electron shape along both longitudinal and transverse directions for slow-electrons. In a system with fixed values of laser parameters, such as the wavelength, the intensity, and the polarization, the gap spacing between two effective nanorods is the only way to control the delay and the initial phase of oscillations at each nanorod. Therefore, the sequential initially out-of-phase near-field action is studied by decreasing the gap distance between two nanorods to 80 nm and keeping the inclined laser pulse and electron parameters constant.

As shown in Figure. 4a, I and b, I, the momentum-space distribution of the electron after passing through the first field resembles that of the in-phase case (Figure. 2I,), where the direction of the electric field of the laser radiation is translated into the rotated asymmetric distribution of the electron wavepacket in the momentum space. As the electron travels outside of the first interaction zone, it encounters the near-field of the second nanorod but with reversed direction of Lorentz force with respect to the in-phase system. Hence, a more complicated electron modulation effect is observed. The non-uniform phase of this oscillating optical mode in the second interaction zone induces an inverse rotational wiggling motion compared to the first zone. Consequently, it leads to the depletion of the transversal momentum states for the loss channels and the population of the higher-order momentum states for the energy-gain channels. Simultaneously, this non-uniform transversal electromagnetic field increases the probability of occupying positive diffraction angles. Figure. 4, demonstrated that both accelerated and decelerated electrons after the first interaction experience acceleration in the next cycle, leading to an increased probability of gain and a decreased probability of loss events (the span of energy spectra is altogether between $-25\text{eV} \leq E \leq 50$ eV ).

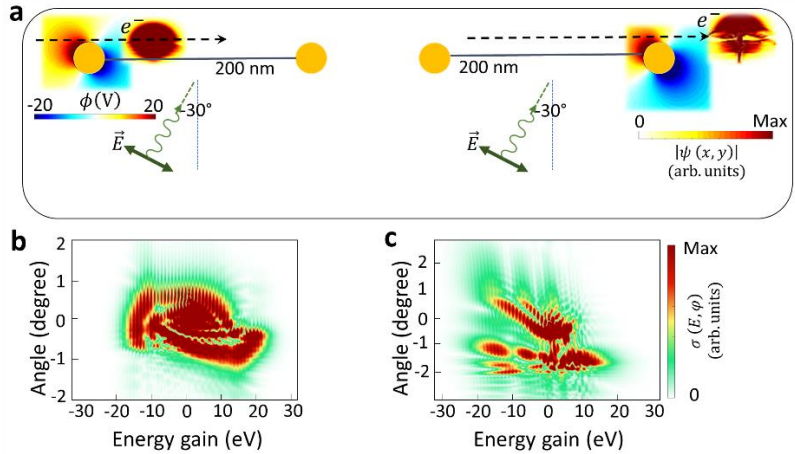

**Figure. 5. Controlling the transverse recoil of the electron as a function of the relative phase between plasmonic gold nanorods.** (a) The amplitude of the electron wavepacket in the real space after the first and second interaction zones. The inset shows the spatial profile of the scalar potentials at given times. The inelastic scattering cross-section of the electron, after its interaction with the excited dipolar near-field of the (b) first, and (c) the second nanorod. Nanorods are excited with pulsed laser beam (laser electric field amplitude, wavelength, and temporal broadening are $E_0 = 1$ GVm$^{-1}$, 700 nm and 18 fs, respectively). The considered electron wavepacket has an initial kinetic energy of 600 eV.

Keeping the electron parameter and laser excitation as before (Figure. 3), the sequential interaction of the electron wavepacket with initially positive optical fields is studied by increasing the gap spacing between gold nanorods (200 nm), in such a way that the experienced electric-field phase by the electron is the same at both interaction zones (Figure. 5). The results support the overall behavior of the smaller gap case shown in Figure. 2. In both cases two distinguished diffraction peaks are observed. This confirms that the major control parameter is the phase offset experienced by the electron beam interacting with both dipolar resonances.

Figure. 5 shows that by adjusting the initial amplitude and phase distribution of the near-fields from an electron perspective and carefully controlling the phase delay between subsequent fields, along with considering a gap spacing smaller than the dispersive length[27] of the electron, we can reach engineered simultaneous quantized energy and transverse momentum exchange between a propagating light wave and a free-electron. Concomitant electron acceleration/deceleration and diffraction caused by sequential in-phase and out-of-phase near-field oscillations allow for designing a highly controllable polarization-dependent electron-beam modulator that enable selective acceleration and deceleration of the electron beams at specific diffraction angles.

**Conclusion**

In conclusion, our work presents a new way to spatially shape electron beams, along not only longitudinal but also transverse directions, by tuning the polarization of the incident light and the distance between two nanorods in a dual-interaction system. In the Ramsey-type method, the direction of the circular wiggling motion in each interaction zone, together with the phase of the oscillating localized plasmon at the starting point of the interaction with the electron, as well as the relative phase between two near fields, can control the energy transfer and recoil experienced by the electron in arbitrary angular deflections. Revealing quantum features in such interactions, we have investigated the role of the initial orientation of the wiggling motion of the electron beam interacting with sequential near-field distributions, thus controlling the overall shape of the electron wavepacket. Our approach provides an efficient way to actively modulate the electron wavefunctions and coherently control the manipulation of free electron waves by tailoring both elastic and inelastic effects, achieving diffraction orders beyond what is normally achievable by the Kapitz-Dirac effect. This finding motivates further studies that considers the development of novel techniques for ultrafast electron-light interferometry and shaping the electron wave packet for further investigations of quantum coherent interactions in complex systems.

**Acknowledgement:**

This project has received funding from the European Research Council (ERC) under the European Union's Horizon 2020 research and innovation program, Grant Agreement No. 802130 (Kiel, NanoBeam) and Grant Agreement No. 101017720 (EBEAM).

**Supplementary Material:**

Supplementary Note 1: Shaping Electron wave packet with Single Gold Nanorod

Supplementary Note 2: Influence of the Initial Optical Phase on the Electron Energy Spectra

Supplementary Note 3: Elastic Transverse Diffraction of Electron from Dual Interaction System

Supplementary Note 4: Influence of the Wavepacket Dimension

**Supplementary Note 1: Shaping Electron wave packet with Single Gold Nanorod**

To fully understand the impact of the considered gap distance between gold nanorods on their hybridization within the dual interaction system, considering the dynamics of the recoil exerted by a single gold nanorod is important. Figure. S1 demonstrates the electron response to the laser-induced plasmonic near-field of a single nanorod when the initial parameters of the electron and laser in the simulation domain are the same as the dual-interaction scheme. This study resembles the electron modulation occurring within the first near-field zone in the sequential interaction for both direction of laser polarization ($\theta = 0°$, and $\theta = -30°$). Hence, for the small nanorods and short electron wave packet, which are the considered parameters in this work, the gap spacing between these two effectives near-field zone does not lead to the plasmonic hybridization effect[1,2].

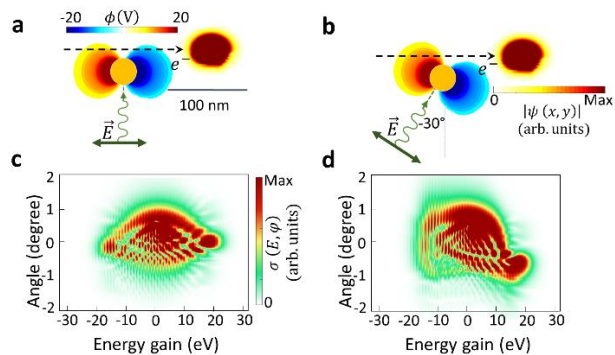

**Figure. S1. Modulation of the electron wavepacket after the interaction with laser-induced dipolar plasmon oscillations in a gold nanorod.** (a) Scalar potential and amplitude of the electron wave function after the interaction with linearly polarized light in (a) normal and (b) oblique excitation scenarios. Inelastic scattering cross-section map corresponding to (c) normal and (d) oblique excitation scenarios, specifying no hybridization in a double interaction system with a 100 nm distance between gold nanorods.

**Supplementary Note 2: Influence of the Initial Optical Phase on the Electron Energy Spectra**

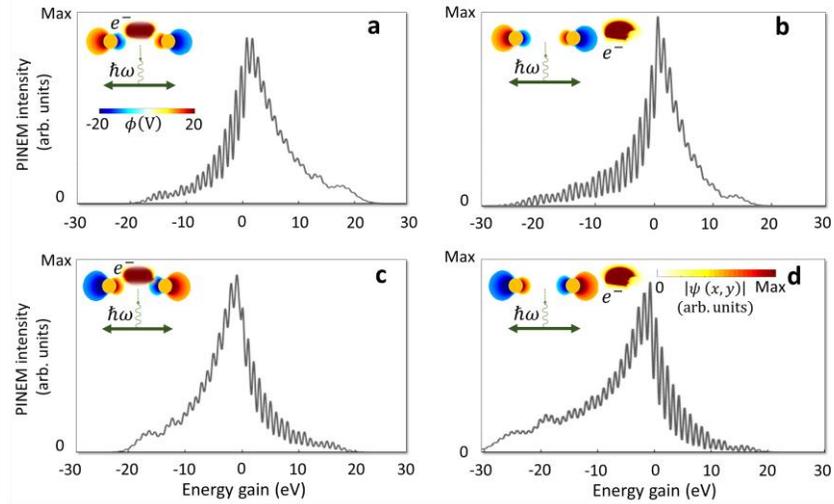

**Figure. S2. PINEM spectra.** Energy gain spectra after the interaction with the system specified as inset. (a) Single and (b) dual interactions with the depicted positive phase of *x*-polarized initial scaler potential. (c) First and (d) second interactions are initialized by the negative phase of potential. Correlation between the entrance phase and initial interaction mode of the system caused selective accelerated and decelerated PINEM maps.

The final longitudinal energy comb for individual interaction highlights the importance of synchronous motions between near-fields oscillations and electron wavepackets. It has already been shown that in a dual interaction system the combination of the coupling strength from each near-field $g_1, g_2$ and the phase delay between two spatially varying fields can control the final energy distribution. Here our finding indicates the correlation between the initial phase of the oscillating plasmon fields in final energy modulation of the electron. The PINEM map shows the exchange of $n$ photons with the energies of $\hbar\omega_{ph}$ within the energy range of $-20\text{eV} \leq E \leq 20$ eV. The dual in-phase interaction system for a weak-interaction regime (intensity of zero-line energy peak is more than inelastic peaks) can control the probability amplitude of each photon order. With this the intensity of the spectral peak for already gained electron is doubled, and for the loss channels is suppressed (Figure. S2). When the initial effective phase of the near-field is positive (Figure. S2 a, b) the probability of energy loss events are more than the gain events. This effect is further enhanced in the second interaction zone. The scenario is reversed when the negative lobe of the scaler potential at two spots starts affecting the propagating electron (Figure. S2 c, d).

# Supplementary Note 3: Elastic Transverse Diffraction of Electron from Dual Interaction System

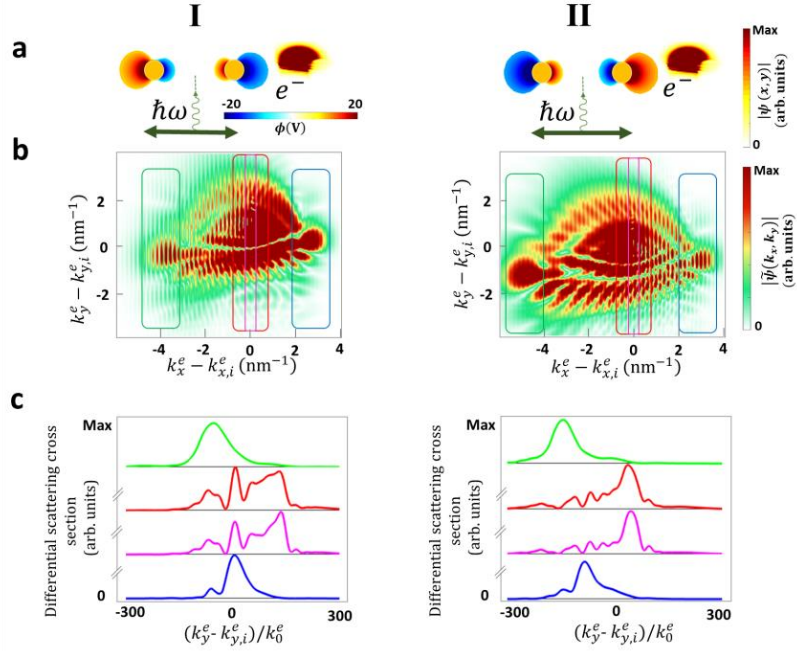

**Figure. S3. Effect of the phase of the near fields on the transverse recoil of free electrons.** The electron initiates its interaction with plasmonic nanorods in a sequential (I) positive, and (II) negative optical phase of the dipolar mode. (a) Schematic of final modulation of the amplitude of the electron wavepacket in real space, and (b) the electron wavepacket in momentum space, after dual interaction with the *x*-polarized dipolar localized plasmons of the gold nanorods. (c) Electron distribution along the transverse direction integrated over the narrow longitudinal ranges shown by the highlighted colored boxes in panel b. The electron has the kinetic energy of 600 eV ($\beta = 0.048$), excited with a laser pulse at the wavelength of 700 nm, and field amplitude of $E_0 = 1$ GVm$^{-1}$.

To enhance our understanding of the physics of sequential electron interaction with near-field light, we studied the differential scattering cross-section (Figure. S3). To do so, the values of the vertical recoil is averaged over the narrow longitudinal energy ranges (corresponding highlighted energy ranges shown in the inelastic scattering cross section map). Within this configuration, the transversal modulation of the electron wavepacket is better visualized by the line profiles.

The studies for two systems involving subsequent positive (Figure. S3, I), and negative (Figure. S3, II) initial phases of the scaler potential demonstrated the critical role of the arrival time of the electron to the near-fields zone. Assigning the impact of the circular wiggling motion to control the diffraction angle and establish the selection roles for transverse recoil with x-polarized light, we can see distinguishable momentum peaks along positive and negative angles.

As demonstrated in column I of Figure. S3, for electron experience significant elastic scattering (pink, and red boxes), the transversal momentum distribution sustains separate peaks in a diffraction angle close to zero, where the highest intensity after averaging is at zero $k_y$. Whereas, for inelastic contributions (green, and blue boxes) there is a sharp diffraction peak broadening close to the zero $k_y$. The diffracted peaks in the other system (column II) show two peaks of highly pronounced negative transverse diffraction for inelastically scattered electrons (green, and blue boxes). While for the elastic contribution (pink, and red boxes) it supports one sharp peak at zero and several weaker resonances at negative values of $k_y$. Indeed, the relative phase between near-fields in the sequential structure adds another degree of freedom to control the transversal recoil and diffraction peaks and is responsible for producing highly separated spectral fringes.

**Supplementary Note 4: Influence of the Wavepacket Dimension**

Here we provide additional discussion on the effect of the longitudinal broadening of the electron wavepackets on the inelastic energy transfer as well as the elastic diffraction of the electron beam. The simulation parameters here are chosen similar to those presented in Figure. 2a and b of the manuscript with a longer electron pulse. Our simulation shows that involving a larger spatial spread of the electron wavepacket (Figure. S4), which has a longer effective interaction length with the rotational field (4 oscillation period for 80 nm, and 6 oscillation period for 120 nm), supports the results for the smaller broadening of the electron wave packet. The comparison indicates constructive interference paths originates from the longer electron pulse can ensure discrete and distinguishable momentum modulation like the shorter electron pulses.

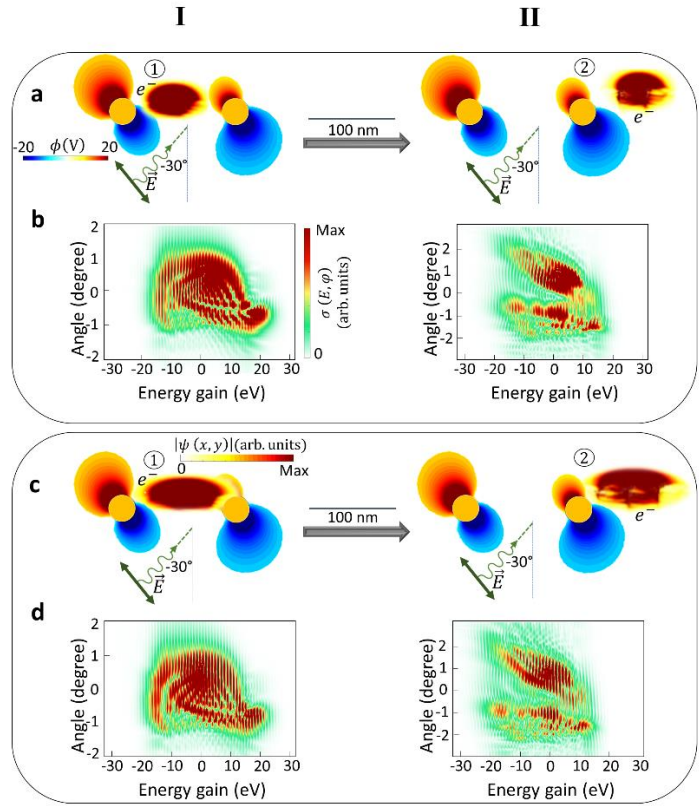

**Figure. S4. Effect of the longitudinal broadening of the electron pulse on the final momentum recoil of the electron wavepacket after the sequential interaction.** Final distribution of the transversal and the longitudinal modulation of free electrons in (a, c) real space, and (b, d) its inelastic scattering cross-section after each process (highlighted with 1 and 2 in the figures). The electron wave packet has an initial kinetic energy of 600 eV, with 36 nm transverse broadenings, 80 nm (upper box), and 120 nm (lower box) longitudinal broadenings, respectively (FWHM). Laser electric field amplitude, wavelength, and temporal broadening are $E_0 = 1 \text{ GVm}^{-1}$, 700 nm and 18 fs, respectively.

Particularly, for the electron with 120 nm longitudinal broadening, the aforementioned transversal electron deflection in real space is more visible for the bunched electron. This work is experimentally achievable, especially when samples are placed at distances more than 200 nm less than the dispersion length.